\documentclass[12pt]{iopart}

\usepackage{iopams,subeqn}  
\usepackage{color}

\usepackage{graphicx}

\begin{document}

\title[Open TASEP with local resetting]{Totally asymmetric simple exclusion process with local resetting and open boundary conditions}

\author{A Pelizzola$^{1,2,3}$, M Pretti$^{3}$}

\address{$^1$ Dipartimento Scienza Applicata e Tecnologia, Politecnico di Torino - Corso Duca degli Abruzzi 24, I-10129 Torino, Italy}
\address{$^2$ INFN, Sezione di Torino - Via Pietro Giuria 1, I-10125 Torino, Italy}
\address{$^3$ Consiglio Nazionale delle Ricerche - Istituto Sistemi Complessi (CNR-ISC) c/o DISAT, Politecnico di Torino - Corso Duca degli Abruzzi 24, I-10129 Torino, Italy}
\ead{alessandro.pelizzola@polito.it, marco.pretti@polito.it}

\vspace{10pt}
\begin{indented}
\item[]October 2022
\end{indented}

\begin{abstract}
  We study a \emph{totally asymmetric simple exclusion process} with open boundary conditions and local resetting at the injection node. 
  We investigate the stationary state of the model, using both mean-field approximation and kinetic Monte Carlo simulations, and identify three regimes, depending on the way the resetting rate scales with the lattice size. 
  The most interesting regime is the intermediate resetting one, as in the case of periodic boundary conditions. 
  In this regime we find pure phases and phase separation phenomena, including a low-density/high-density phase separation, which was not possible with periodic boundary conditions. 
  We discuss density profiles, characterizing bulk regions and boundary layers, and nearest-neighbour covariances, finding a remarkable agreement between mean-field and simulation results. 
  The stationary state phase diagram is mapped out analytically at the mean-field level, but we conjecture that it may be exact in the thermodynamic limit. 
  We also briefly discuss the large resetting regime, which exhibits an inverse characteristic length scale diverging logarithmically with the lattice size. 
\end{abstract}

%
%
%
\maketitle
%
%

\section{Introduction}

Simple exclusion processes are Markov stochastic processes of fundamental importance in non-equilibrium statistical physics.
They are highly simplified models whose behaviour, especially in the stationary state, can be extremely rich (for recent reviews see \cite{Derrida98,Schutz,Evans2007,ZiaReview,TransportBook}). 
A prominent example is the \emph{totally asymmetric simple exclusion process} (TASEP), in which particles diffuse on a one-dimensional lattice, subject to a bulk drive (particles hopping is allowed only in one direction, say rightward to fix ideas) and to an exclusion constraint (at most one particle can occupy a lattice node). 
In case of open boundary conditions (OBCs) a boundary drive is also present (particles can enter the lattice at the left boundary and leave at the right boundary), and the model behaviour is much richer than in case of periodic boundary conditions (PBCs). 
Many exact results are available for this basic model, and various generalizations and extensions have been proposed, typically studied by approximate methods, such as mean-field (MF) approximations, and by kinetic Monte Carlo (KMC) simulations. 

One generalization of the basic TASEP with OBCs which is particularly relevant for the present work is the totally asymmetric simple exclusion process with \emph{Langmuir kinetics} (TASEP-LK) \cite{ParmeggianiFranoschFrey03,Popkov,Evans2003,ParmeggianiFranoschFrey04}. 
In the latter model, particles can also attach to an empty node with rate $\omega_\mathrm{A}$ or detach from an occupied one with rate $\omega_\mathrm{D}$. 
These attachment and detachment processes can play a non-trivial role, provided the corresponding rates scale appropriately with the lattice size $L$ in the thermodynamic limit ${L \to \infty}$. 
In particular, due to the bulk nature of these processes, it can be easily understood that their rates $\omega_\mathrm{A,D}$ must scale as $L^{-1}$, so that the ``macroscopic rates'' $\Omega_\mathrm{A,D} = L \omega_\mathrm{A,D}$ remain finite for $L \to \infty$.
The most relevant and known effect of Langmuir kinetics, in the aforementioned regime, is the onset of phase coexistences, with the density profile characterized by shocks (or domain walls) between regions at different densities, which remain localized and stable over time.
Similar shocks have also been predicted in models characterized by a point defect (i.e.~a slower hopping rate)~\cite{JanowskyLebowitz1992} or more generally by non-uniform hopping rates~\cite{BanerjeeBasu2020}.
TASEP-like models have also been studied, in which the interplay between both mechanisms, Langmuir kinetics and defects, gives rise to a very rich phenomenology~\cite{PierobonMobiliaKouyosFrey2006}. 

Another, very recent, generalization, which will be considered here, was introduced in \cite{Reuveni} and further investigated in \cite{EPL21}. 
Reference \cite{Reuveni} considers a \emph{symmetric simple exclusion process} with \emph{local resetting} (SSEP-LR) and PBCs. 
Local resetting, where particles can reset their position independently of one another, is more challenging than the global resetting considered in previous works \cite{Pal,Nagar}, where the whole system is simultaneously reset to some reference state. 
In particular, the approach based on renewal theory (see \cite{Majumdar} and references therein, also for a general perspective on stochastic resetting) cannot be applied in the case of local resetting. 
In \cite{EPL21} it was shown that the behaviour of the SSEP-LR in the thermodynamic limit depends crucially on the way the resetting rate $r$ scales with $L$, and that this behaviour is especially rich in an intermediate resetting regime where $r \sim L^{-2}$. 
The analysis was then extended to the TASEP with local resetting (TASEP-LR) and PBCs, showing that the intermediate resetting regime arises for $r \sim L^{-1}$, and pointing out a relationship between this model and the TASEP-LK, in the special case in which only the detachment process is present (from the bulk of the system). 
In both works \cite{Reuveni,EPL21} a remarkable agreement between MF and KMC results was found.

In the present paper we proceed along this line of investigation, considering a TASEP-LR with OBCs and resetting at the injection node, and studying the stationary state of the model using both MF and KMC.
The subject of this work is thus a model system, designed in order to investigate the role of the local resetting mechanism in an interacting particle system such as the open TASEP, of very general interest in non-equilibrium statistical physics. 
It is worth mentioning that the resulting model has some interesting, at least qualitative, analogies with certain models of biological microsystems. 
First of all, let us remember that the TASEP itself was originally conceived~\cite{MacDonaldGibbsPipkin1968} as a model for the dynamics of ribosomes on the mRNA (polyribosome), at the core of the protein synthesis process. 
In this framework, particles represent ribosomes, whereas lattice nodes represent codons, i.e.~the basic information units of mRNA. 
Furthermore, the ``ordinary'' TASEP-LK (with both attachment and detachment processes) has been considered as a minimal model for the dynamics of molecular motors on microtubules, i.e.~for intracellular transport processes~\cite{FreyParmeggianiFranosch2004,AppertrollandEbbinghausSanten2015}. 
Finally, the TASEP-LK with only detachment kinetics has been recently studied~\cite{BonninKernYoungStansfieldRomano2017}, as a refined version of TASEP, in order to take into account, still in the context of the polyribosome, the so-called \emph{drop-off} phenomenon (also known as \emph{nonsense} or \emph{abortion}), i.e.~the premature termination of the translation process, due to detachment of a ribosome before the stop codon.
In this context, our model with local resetting may represent a mechanism of ribosome rescue and recycling~\cite{FranckenbergBeckerBeckmann2012}, with the resetting node representing the mRNA position where ribosomes bind in order to start translation (the so-called Shine-Dalgarno sequence for prokaryotes~\cite{BiologyBook}). 
Of course, this correspondence is to be seen in an effective sense, since it is known that in general ribosome recycling is a complex process, which also requires disassembly of each ribosome into sub-units, followed by diffusion in the cellular environment. 

The plan of the paper is as follows: in section \ref{sec:model} we describe the model and the MF approximation, impose the stationary state conditions and take a continuum limit; in section \ref{sec:results} we present our results, focusing mainly on the intermediate resetting regime; in section \ref{sec:discuss} we draw our conclusions and outline possible future developments.

\section{Model and mean-field approximation}
\label{sec:model}

\begin{figure}
	\centerline{\includegraphics[width=0.75\textwidth,bb=0 420 450 720]{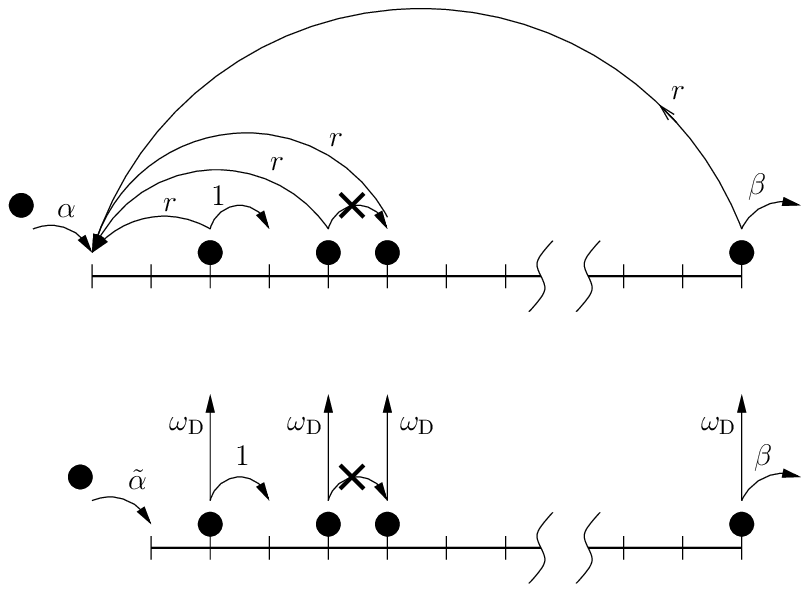}}
	\caption{Kinetic schemes of TASEP-LR (top) and TASEP-LK with detachment only (bottom). 
			It is assumed that lattice nodes are ordered from left to right, and the rate of each process is denoted by a corresponding symbol. 
			The hopping process has unit rate if the arrival node is empty, or it is forbidden (zero rate) otherwise.}
	\label{fig:models}
\end{figure}

We consider the TASEP, with local resetting at the injection node, on a one-dimensional lattice with open boundaries. 
The lattice has $L$ nodes, and a time-dependent occupation number $n_l^t$ is associated to each node ${l = 1,\ldots, L}$. 
We define ${n_l^t = 1}$ (respectively 0) if node $l$ is occupied by a particle (resp.\ empty) at (continuous) time $t$. 
A particle at node $l$ can hop to node $l+1$ with unit rate, provided the destination node is empty. 
Particles are injected at node 1 (if empty) with rate $\alpha$, and extracted from node $L$ (if occupied) with rate $\beta$. 
In addition to these processes, characterizing the ordinary open TASEP, we have the local resetting process: a particle at a node $l > 1$ can hop to node 1 (the injection node), with rate $r$, as usual provided the destination node is empty.
A scheme of all the elementary processes, defining the TASEP-LR model, is reported in figure~\ref{fig:models}. 

In order to apply the MF approximation, we define the local densities $\rho_l^t = \langle n_l^t \rangle$ and we recall that MF amounts to neglecting correlations, thereby approximating $\langle n_k^t n_l^t \rangle \simeq \langle n_k^t \rangle\langle n_l^t \rangle = \rho_k^t \rho_l^t$. 
Following \cite{EPL21}, in particular equations (12)-(13), and replacing periodic with OBCs, we obtain MF equations for the time evolution of the local densities:
\begin{subequations}
\begin{eqnarray}
	\dot \rho_1^t &=& \alpha (1 - \rho_1^t) - \rho_1^t (1 - \rho_2^t) + r (1 - \rho_1^t) \sum_{l=2}^{L} \rho_l^t 
	\, , \\
	\dot \rho_l^t &=& \rho_{l-1}^t (1 - \rho_l^t) - \rho_l^t (1 - \rho_{l+1}^t) - r (1 - \rho_1^t) \rho_l^t 
	\quad \ (l = 2, \ldots, L-1) 
	\, , \\
	\dot \rho_L^t &=& \rho_{L-1}^t (1 - \rho_L^t) - \beta \rho_L^t - r (1 - \rho_1^t) \rho_L^t 
	\, .
\end{eqnarray}
\end{subequations}
In the following we shall focus on the stationary state, where the above differential equations reduce to a set of $L$ algebraic equations for the stationary densities. 
Dropping the time index, we denote by $\rho_l$ the local density at node $l$ in the stationary state and obtain
\begin{subequations}
\begin{eqnarray}
	0 &=& \alpha (1 - \rho_1) - \rho_1 (1 - \rho_2) + r (1 - \rho_1) \sum_{l=2}^{L} \rho_l 
	\, , \\
	0 &=& \rho_{l-1} (1 - \rho_l) - \rho_l (1 - \rho_{l+1}) - r (1 - \rho_1) \rho_l \qquad (l = 2, \ldots, L-1) 
	\label{eq:NESSbulk} 
	\, , \\
	0 &=& \rho_{L-1} (1 - \rho_L) - \beta \rho_L - r (1 - \rho_1) \rho_L
	\label{eq:NESSL}
	\, .
\end{eqnarray}
\end{subequations}
These equations are easily and efficiently solved as in \cite{Botto2019}, by rewriting them in the fixed point form 
\begin{subequations}
\begin{eqnarray}
	\rho_1 &=& \frac{\alpha + r \sum_{l=2}^{L} \rho_l}{1 - \rho_2 + \alpha + r \sum_{l=2}^{L} \rho_l} 
	\label{eq:NESS1-FP} 
	\, , \\
	\rho_l &=& \frac{\rho_{l-1}}{1 - \rho_{l+1} + \rho_{l-1} + r (1 - \rho_1)} \qquad (l = 2, \ldots, L-1) 
	\label{eq:NESSbulk-FP} 
	\, , \\
	\rho_L &=& \frac{\rho_{L-1}}{\beta + \rho_{L-1} + r (1 - \rho_1)} 
	\label{eq:NESSL-FP}
	\, ,
\end{eqnarray}
\end{subequations}
which turns out to always converge to a solution.

As in \cite{EPL21} we observe that a relationship with TASEP-LK can be found, at least in the stationary state and at the MF level. 
In particular, equations (\ref{eq:NESSbulk}) and (\ref{eq:NESSL}) are equivalent to the MF equations for the stationary state of a TASEP-LK with ${L-1}$ nodes, OBCs, injection rate ${\tilde{\alpha} = \rho_1}$ at node 2, extraction rate $\beta$ at node $L$, attachment rate ${\omega_\mathrm{A} = 0}$ and detachment rate ${\omega_\mathrm{D} = r \left( 1 - \rho_1 \right)}$. 
To clarify the relationship, in figure~\ref{fig:models} we have also reported a scheme of the processes involved in the TASEP-LK model (depicted without node~1 and in the special case of detachment-only Langmuir kinetics). 
In particular, one can argue that the two models are not trivially equivalent, in that the effective parameters $\omega_\mathrm{D}$ and $\tilde{\alpha}$ of the TASEP-LK model should in principle depend on the occupation number of node~1, which is a dynamical variable for TASEP-LR. 
As mentioned above, a precise relationship holds only at the MF level, where one can replace the occupation number with the occupation probability, and in the stationary state, where the latter no longer depends on time. 

Equation (\ref{eq:NESSbulk}), describing the bulk behaviour, is the same as in the PBCs case \cite{EPL21}, and so is its continuum (hydrodynamic) limit, which we recall here. 
Assuming ${L \gg 1}$ and defining the scaled position variable ${x = l/L \in (0,1]}$, the macroscopic resetting rate ${R = r L}$ and the parameter  
\begin{equation} 
\lambda = R \left( 1 - \rho_1 \right)
\, , 
\label{eq:definizione_lambda}
\end{equation}
we obtain a differential equation for the function $\rho(x)$, which we shall denote as bulk (or continuum) density profile: 
\begin{equation}
\frac{\rmd \phantom{x}}{\rmd x} 
\left[ \rho(x) \left( 1 - \rho(x) \right) \right] 
= - \lambda \rho(x) 
\, .
\end{equation}
In terms of the composite function $F(\rho(x))$, where ${F(\rho) = \rho \rme^{- 2 \rho}}$, we then have
\begin{equation}
\frac{\rmd \phantom{x}}{\rmd x} F(\rho(x)) 
= - \lambda F(\rho(x)) 
\, ,
\label{eq:fxdiff}
\end{equation}
where $\lambda$ clearly plays the role of an inverse length scale. 
The macroscopic resetting rate $R$ is a key parameter, whose behaviour in the thermodynamic limit ${L \to \infty}$ determines the relevance of the resetting process, relative to ordinary TASEP. 
The most interesting situation is obtained when $R$ tends to a positive constant, that is ${r \sim 1/L}$. 
In this case we speak of intermediate resetting, and in the following we shall focus mainly on this case, discussing briefly the cases ${R \to 0}$ and ${R \to \infty}$, which we shall refer to as small and large resetting, respectively. 
In the intermediate and large resetting cases we shall use the solution of equation (\ref{eq:fxdiff}) in the form
\begin{equation}
  F(\rho(x)) = 
  F(\rho(x_0)) \, \rme^{- \lambda (x-x_0)},
  \label{eq:fx}
\end{equation}
where $x_0$ is some reference point. 
In particular it will be convenient to take as reference points ${x_0 = 0,1}$ (namely, the left or right boundary), so we introduce specific symbols for the corresponding values of $\rho(x)$, namely ${\rho_+ = \rho(0)}$ and ${\rho_- = \rho(1)}$. 
Note that in general we may have ${\rho_+ \neq \rho_1}$ and/or ${\rho_- \neq \rho_L}$ (even in the thermodynamic limit), due to the onset of so-called boundary layers. 
In pure phases (that is, in the absence of shocks in the density profile), from equation \eref{eq:fx} we immediately get that $\rho_+$ and $\rho_-$ are related by
\begin{equation}
F(\rho_-) = F(\rho_+) \, \rme^{-\lambda} 
\, .
\label{eq:relazione_bordi}
\end{equation}
In order to solve the latter equation (for either $\rho_+$ or $\rho_-$), and equation \eref{eq:fx} for the whole bulk profile $\rho(x)$, it is useful to recall that the inverse of $F(\rho)$ is a multivalued function with 2 real branches (corresponding to low and high density phases, respectively), related to the Lambert $W$ function as 
\begin{equation}
{F}^{-1}(\phi) =  
\cases{
	- {\textstyle \frac{1}{2}} W_{ 0}(-2\phi) & $\leq \frac{1}{2}$  \\  
	- {\textstyle \frac{1}{2}} W_{-1}(-2\phi) & $\geq \frac{1}{2}$
}
\, ,
\label{eq:funzione_inversa}
\end{equation}
where $W_{0}$ and $W_{-1}$ denote the 2 real branches of the Lambert function. 
The right boundary value $\rho_-$ enters also a balance equation, relating the injection rate $\alpha$ and the local density $\rho_1$ at the injection node. 
In the stationary state, the injection current ${\alpha (1-\rho_1)}$ must obviously equal the extraction current ${\beta \rho_L}$, whereas the latter can be expected to equal the hopping current close to the extraction node, provided the resetting current stemming from a microscopic layer is negligible in the thermodynamic limit. 
As a consequence we can write  
\begin{equation}
\alpha (1-\rho_1) = \rho_- (1-\rho_-) 
\, .
\label{eq:kirchhoff}
\end{equation}
Note that this last equation is crucial to distinguish the OBCs from the PBCs case.

\section{Results}
\label{sec:results}

In the present section we discuss the stationary state, using both the continuum limit of the MF approximation, discussed in the previous section, and finite size results from KMC simulations, carried out using Gillespie's algorithm. 
Simulations running time is $10^7$, and averages are taken in the stationary state, for ${t \in [10^6,10^7]}$. 
Most of the section deals with the intermediate resetting regime (specifically we fix ${R = 0.5}$, unless otherwise stated), while small and large resetting regimes are discussed in the last subsection.

\subsection{Pure phases} 

\begin{figure}
	\centerline{\includegraphics[width=0.7\textwidth]{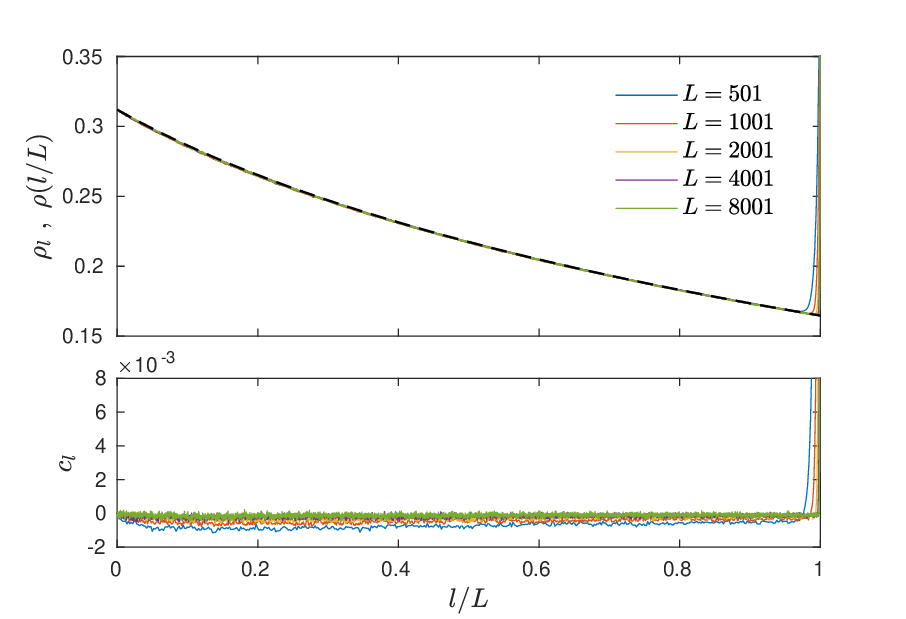}}
	\caption{Stationary density profile (top panel) and covariances (bottom panel) in the LD phase: ${\alpha = 0.2}$, ${\beta = 0.3}$.
	Colored lines denote KMC results.
	A black dashed line represents the MF continuum density profile, defined by equation \eref{eq:fase_LD_prof}.}
	\label{fig:LD_density}
\end{figure}

At small injection rate $\alpha$ and large extraction rate $\beta$ the stationary state is in a low-density (LD) phase. 
Like in the PBCs case \cite{EPL21} the LD phase is characterized, at the MF level, by
\begin{equation}
\rho_+ = \rho_1 
\, ,
\label{eq:fase_LD}
\end{equation}
expressing the absence of a left boundary layer.
Given the model parameters $\alpha$, $\beta$ and $R$, the latter equation together with \eref{eq:definizione_lambda}, \eref{eq:relazione_bordi} and \eref{eq:kirchhoff} allow us to determine $\rho_+$, $\rho_-$, $\rho_1$ and $\lambda$, and hence, by 
\eref{eq:fx} with ${x_0=0}$ and \eref{eq:funzione_inversa}, the continuum density profile
\begin{equation}
\rho(x)
= - \textstyle{\frac{1}{2}} W_0 \left( -2 F(\rho_1) \, \rme^{-\lambda x} \right) 
\, .
\label{eq:fase_LD_prof}
\end{equation}
In figure \ref{fig:LD_density} (top panel) we report the above profile, along with stationary density profiles obtained by KMC simulations, for 5 different lattice sizes $L$, at ${\alpha = 0.2}$ and ${\beta = 0.3}$. 
The collapse of KMC data confirms that the density profile depends on resetting rate $r$ and system size $L$ only through ${R = rL}$, as predicted by the MF theory, which agrees very well with the numerical results. 
In the bulk and at the left boundary (close to the injection node) finite size effects are not appreciable on the drawing scale, and KMC results practically coincide with MF ones already at the smallest size considered. 
This is confirmed by the behaviour of the nearest-neighbour (NN) covariances ${c_l = \langle n_l n_{l+1} \rangle - \langle n_l \rangle \langle n_{l+1} \rangle}$ from KMC simulations (bottom panel), which in these regions seem to vanish in the thermodynamic limit. 
Let us note that in the LD phase the bulk density profile is determined only by $R$ and $\alpha$ (through $\rho_1$), with no dependence on $\beta$ (like in pure TASEP, where it depends only on $\alpha$). 
At the MF level, indeed, one can see that $\beta$ never appears in the equations leading to \eref{eq:fase_LD_prof}. 

\begin{figure}
	\centerline{\includegraphics[width=0.7\textwidth]{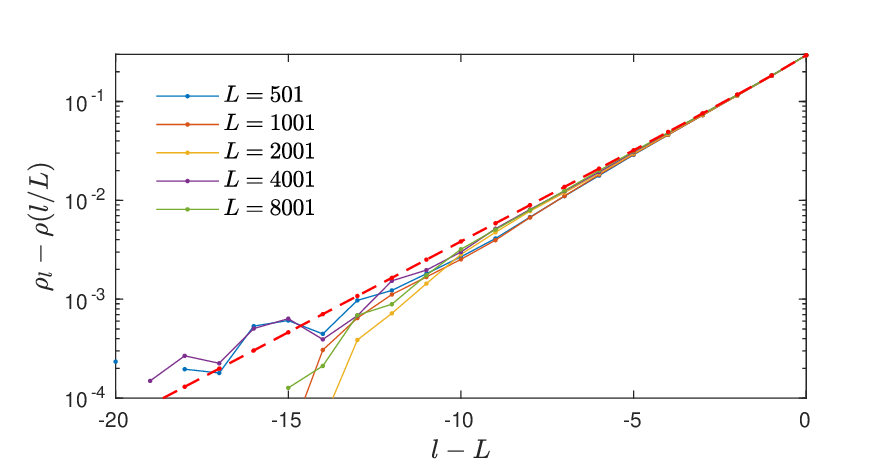}}
	\caption{Right boundary layer of the LD stationary density profile in figure \ref{fig:LD_density}. 
	Solid colored lines denote KMC results. 
	A red dashed line represents the exact boundary layer of an ``equivalent'' pure TASEP in the thermodynamic limit (see the text for details).}
	\label{fig:LD_boundary_log}
\end{figure}

At the right boundary, close to the extraction node, a boundary layer sets on, whose size is microscopic (a few lattice nodes) in the thermodynamic limit, as in the PBCs case \cite{EPL21} (and also in pure TASEP and TASEP-LK \cite{Botto2019}). 
This is the only region where the NN covariances do not vanish in the thermodynamic limit. 
In order to analyze this boundary layer, in figure \ref{fig:LD_boundary_log} we plot the deviation of the KMC density profile with respect to the corresponding bulk profile \eref{eq:fase_LD_prof}, as a function of the node position relative to the right boundary. 
In log-linear scale it is clearly seen that the boundary layer decays exponentially on a scale which is microscopic and independent of the system size (notice that in this case the node position is not scaled by the system size). 
Guided by a similar result for TASEP-LK \cite{Botto2019}, we expect that, in the thermodynamic limit, the boundary layer approaches that of an ``equivalent'' pure TASEP, namely one with injection rate $\rho_-$ (the bulk density value in the vicinity of the boundary layer) and extraction rate $\beta$. 
This is actually what appears from figure \ref{fig:LD_boundary_log}, where the red dashed line denotes the difference between the exact density profile of the pure TASEP (in the thermodynamic limit) and the corresponding bulk density. 

\begin{figure}
	\centerline{\includegraphics[width=0.7\textwidth]{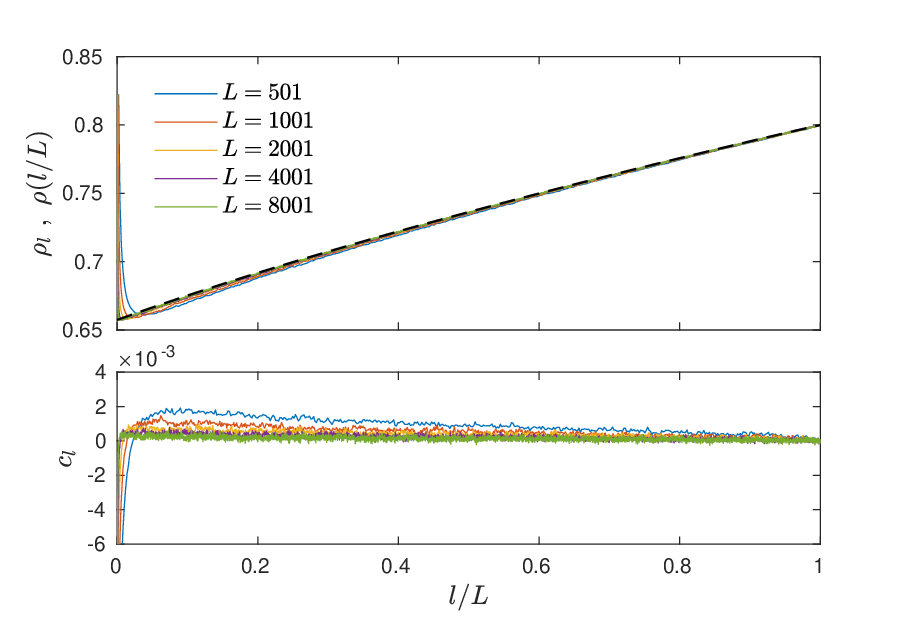}}
	\caption{Same as figure \ref{fig:LD_density} in the HD phase: ${\alpha = 0.9}$, ${\beta = 0.2}$.
	The black dashed line (MF continuum density profile) is defined by equation \eref{eq:fase_HD_prof}.}
	\label{fig:HD_density}
\end{figure}

At large injection rate $\alpha$ and small extraction rate $\beta$ the stationary state is in a high-density (HD) phase. The HD phase is characterized, at the MF level, by
\begin{equation}
\rho_- = 1-\beta 
\, ,
\label{eq:fase_HD}
\end{equation}
expressing the absence of a right boundary layer.
As for the LD phase, the above equation together with \eref{eq:definizione_lambda}, \eref{eq:relazione_bordi} and \eref{eq:kirchhoff} allow us to determine $\rho_+$, $\rho_-$, $\rho_1$ and $\lambda$, and hence, still by 
\eref{eq:fx} but with ${x_0=1}$, the bulk density profile
\begin{equation}
\rho(x)
= - {\textstyle \frac{1}{2}} W_{-1} \left( -2 F(1-\beta) \, \rme^{\lambda (1-x)} \right) 
\, .
\label{eq:fase_HD_prof}
\end{equation}
In figure \ref{fig:HD_density} we report the above profile, together with density and NN-covariance profiles from KMC simulations, still for 5 different lattice sizes, at ${\alpha = 0.9}$ and ${\beta = 0.2}$. 
Similar considerations apply as in the case of the LD phase, but the boundary layer is now at the left boundary, close to the injection node. 
Let us note that in the HD phase the bulk density profile depends on the full set of model parameters $R$, $\alpha$ and $\beta$ (at odds with pure TASEP, where it depends only on $\beta$). 
This can also be argued from the MF equations leading to \eref{eq:fase_HD_prof}. 

\begin{figure}
	\centerline{\includegraphics[width=0.7\textwidth]{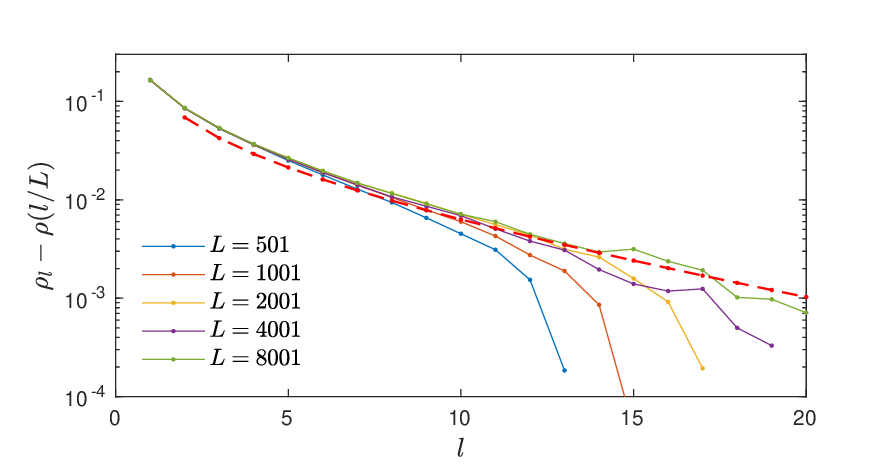}}
	\caption{Same as figure \ref{fig:LD_boundary_log} for the left boundary layer of the HD phase: ${\alpha = 0.9}$, ${\beta = 0.2}$, bulk profile $\rho(x)$ defined by \eref{eq:fase_HD_prof}. 
	Note that the effective pure TASEP starts at node ${l = 2}$.} 
	\label{fig:HD_boundary_log}
\end{figure}

\begin{figure}
	\centerline{\includegraphics[width=0.7\textwidth]{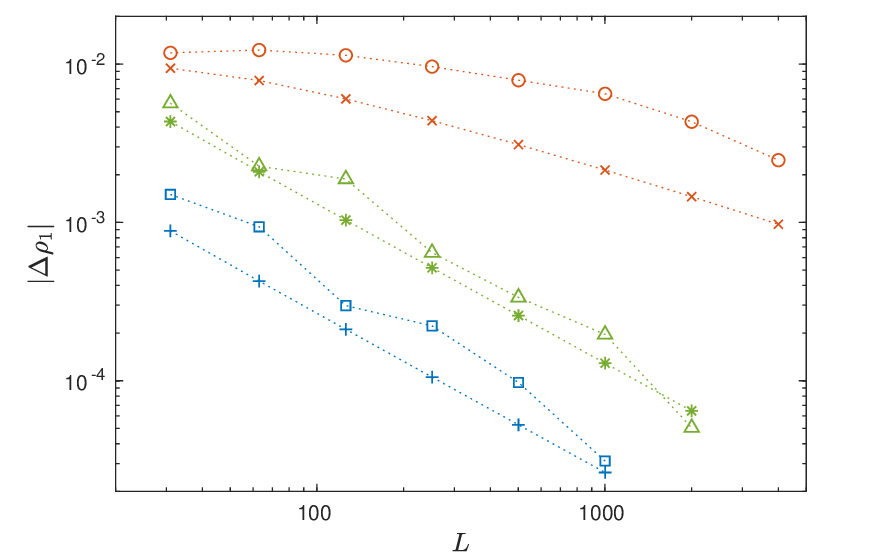}}
	\caption{Absolute difference $\left|\Delta\rho_1\right|$ as a function of the lattice size $L$ (see the text). \\ 
		HD phase (${\alpha = 0.9}$, ${\beta = 0.2}$): blue squares (KMC) and $+$ symbols (MF). \\
		MC phase (${\alpha = 0.4}$, ${\beta = 0.3}$): red circles (KMC) and $\times$ symbols (MF). \\ 
		MC-HD coexistence (${\alpha = 0.3}$, ${\beta = 0.15}$): green triangles (KMC) and $*$ symbols (MF).}
	\label{fig:rho1}
\end{figure}

The detailed analysis of the boundary layer, reported in figure \ref{fig:HD_boundary_log}, shows features similar to the LD case, except the fact that it no longer tends to the boundary layer of the corresponding effective pure TASEP (defined with injection rate $\rho_1$ at node $2$ and extraction rate ${1 - \rho_+}$). 
This can be rationalized by observing that the resetting process affects a right boundary layer in a negligible way, due to scaling ${r \sim 1/L}$ and the microscopic size of the layer. 
Conversely, a left boundary layer is subject to the effect of the total resetting current at node $1$, which tends to a finite value in the thermodynamic limit.
Of course, also the MF theory is unable to provide a quantitative prediction for the boundary layer, but interestingly, considering only the injection node (${l=1}$), we have noticed that the density value $\rho_1$ seems to tend, in the thermodynamic limit, to the corresponding (analytical) MF result, discussed in the previous section. 
In figure \ref{fig:rho1} we plot, as a function of the lattice size $L$, the difference $\Delta \rho_1$ (in absolute value) between the injection node density, evaluated by KMC and finite size MF, and the corresponding limit value predicted by the MF theory. 
We can see a power law decay, with the KMC results following quite closely the MF predictions. 
We anticipate that, as already evident from figure \ref{fig:rho1}, this kind of behaviour seems to be confirmed in any phase characterized by a left boundary layer. 
This result is particularly interesting because, as we shall see below (and at odds with ``ordinary'' TASEP-like models without resetting), in the current model the phase diagram is determined not only by bulk densities but also by the density of the single resetting node. 
As a consequence, we are led to believe that the MF phase diagram, derived in subsection \ref{subsec:phd}, should be exact in the thermodynamic limit. 

\begin{figure}
	\centerline{\includegraphics[width=0.7\textwidth]{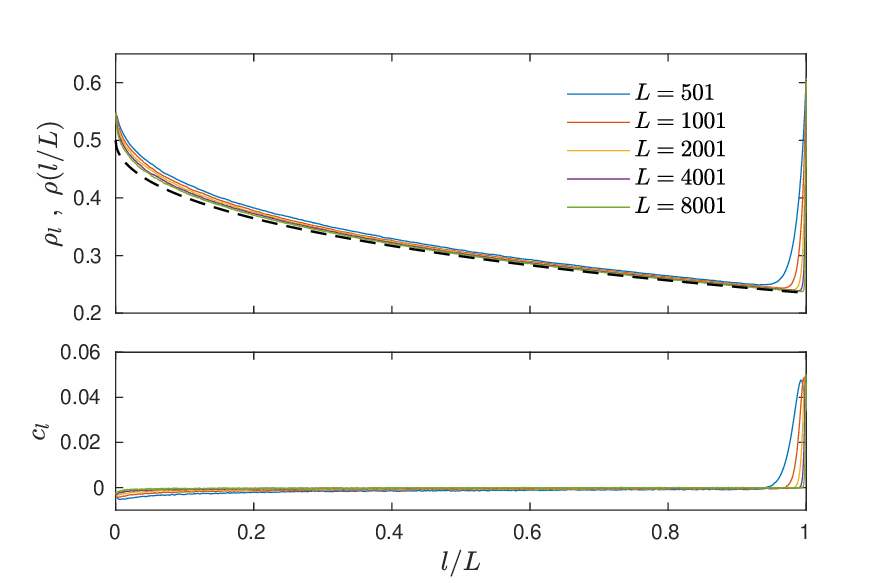}}
	\caption{Same as figure \ref{fig:LD_density} in the MC phase: ${\alpha = 0.4}$, ${\beta = 0.3}$.
	The black dashed line (MF continuum density profile) is defined by equation \eref{eq:fase_MC_prof}.}
	\label{fig:MC_density}
\end{figure}

\begin{figure}
	\centerline{\includegraphics[width=0.7\textwidth]{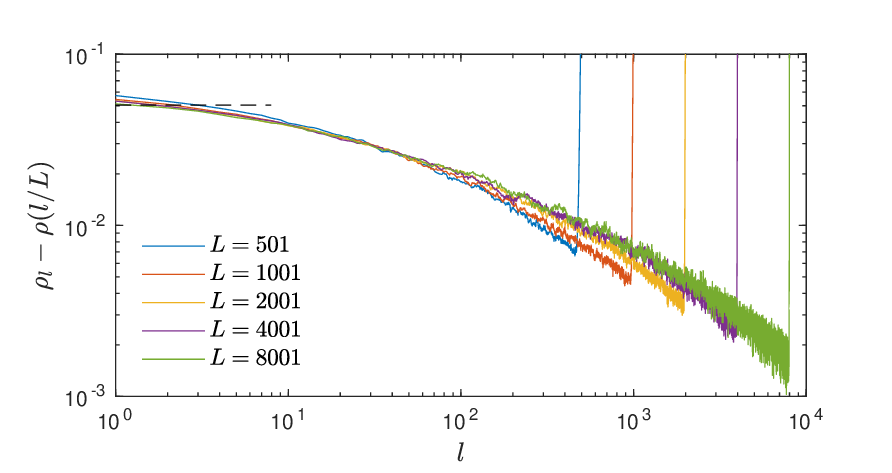}}
	\caption{Same as figure \ref{fig:LD_boundary_log} for the left boundary layer of the MC phase: ${\alpha = 0.4}$, ${\beta = 0.3}$, bulk profile $\rho(x)$ defined by \eref{eq:fase_MC_prof}. 
	A black dashed line marks the (analytical) MF value for ${\rho_1-\rho_+ = \rho_1-1/2}$ in the thermodynamic limit.}
	\label{fig:MC_boundary-sx_log}
\end{figure}

When both the injection rate $\alpha$ and the extraction rate $\beta$ are large enough, we obtain the last pure phase, namely the maximal current (MC) phase. 
At the MF level, this phase is characterized by  
\begin{equation}
\rho_+ = {\textstyle \frac{1}{2}} 
\, ,
\label{eq:fase_MC} 
\end{equation}
which entails that the current close to the injection node, ${J_+  = \rho_+ (1 - \rho_+) = 1/4}$, is maximal. 
As $\rho_+$ is fixed, equations \eref{eq:definizione_lambda}, \eref{eq:relazione_bordi} and \eref{eq:kirchhoff} allow us to determine $\rho_-$, $\rho_1$ and $\lambda$, whereas the continuum MF profile $\rho(x)$ can be determined precisely as in the LD case, by \eref{eq:fx} with ${x_0=0}$ and \eref{eq:funzione_inversa}. 
In the end we obtain 
\begin{equation}
\rho(x)
= - \textstyle{\frac{1}{2}} W_0 \left( -2 F(\textstyle{\frac{1}{2}}) \, \rme^{-\lambda x} \right) 
\, ,
\label{eq:fase_MC_prof}
\end{equation}
analogous to the PBCs case. 
The stationary density profile, displayed in figure \ref{fig:MC_density} (top panel) for ${\alpha = 0.4}$ and ${\beta = 0.3}$, exhibits a bulk region where the KMC results appear again to tend, in the thermodynamic limit, to the MF result, and 2 boundary layers. 
Again, the NN covariances (figure \ref{fig:MC_density}, bottom panel) seem to tend to nonvanishing values only in the boundary layers. 
As in the LD phase, the bulk density profile is determined by $R$ and $\alpha$ (through $\rho_1$), with no dependence on $\beta$ (at odds with pure TASEP, where it is independent of both $\alpha$ and $\beta$).  
The right boundary layer is also analogous to the LD one, and a detailed analysis like that in figure \ref{fig:LD_boundary_log} shows that it tends, in the thermodynamic limit, to that of an equivalent pure TASEP with injection rate $\rho_-$ and extraction rate $\beta$. 
Conversely, the left boundary layer exhibits a decay being clearly much slower than in the LD and HD phases (see figure \ref{fig:MC_boundary-sx_log}), probably characterized by a combination of a power law and an exponential. 
As previously shown in figure \ref{fig:rho1}, the value of ${\rho_1 > \rho_+ = 1/2}$ still seems to tend to the MF result in the thermodynamic limit, as in the HD phase, even though with an apparently slower decay. 

It is worth observing that the dependence of the bulk density profile of the MC phase on the injection rate $\alpha$ is at odds with both pure TASEP and TASEP-LK, and can be specifically ascribed to the introduction of the resetting mechanism.
The injection rate clearly acts on the occupation probability $\rho_1$, while the latter in turn has a regulating effect on the output current of particles from the bulk of the system. 
This last effect controls the non-uniformity of the density profile, even in the MC phase, through the inverse length scale $\lambda$, as one can argue from equation \eref{eq:fase_MC_prof}.

\subsection{Phase separation}

\begin{figure}
	\centerline{\includegraphics[width=0.7\textwidth]{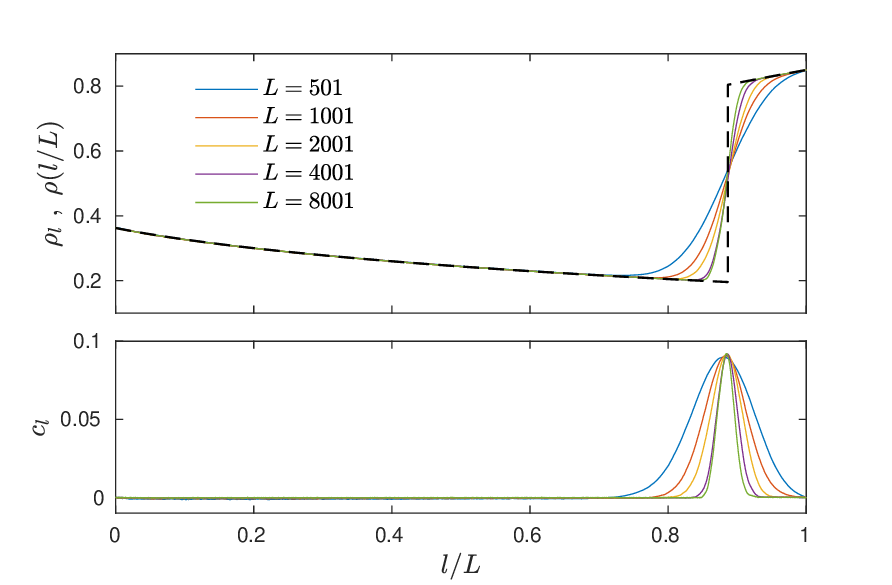}}
	\caption{Same as figure \ref{fig:LD_density} in LD-HD coexistence: ${\alpha = 0.2}$, ${\beta = 0.15}$.
	The black dashed line (MF continuum density profile) is defined piecewise by equations \eref{eq:fase_LD_prof} and \eref{eq:fase_HD_prof}, with the shock position determined by \eref{eq:LD-HD}.}
	\label{fig:LD+HD_density}
\end{figure}

At small injection rate $\alpha$ and intermediate extraction rate $\beta$ we can find LD-HD coexistence, a feature which was not observed in the PBCs case \cite{EPL21}. 
The stationary density profile, illustrated in figure \ref{fig:LD+HD_density} (top panel) for ${\alpha = 0.2}$, ${\beta = 0.15}$, exhibits a LD portion on the left and a HD portion on the right, separated by a domain wall, or shock, without boundary layers. 
Covariances (figure \ref{fig:LD+HD_density}, bottom panel) tend to zero everywhere except at the domain wall, which shrinks as the lattice size $L$ increases. 
Note that in TASEP-LK the domain wall width is known to vanish asymptotically as $L^{-1/2}$ \cite{ParmeggianiFranoschFrey03}, which roughly seems to be the case likewise in the current model.   
At the MF level, the density profile in the LD-HD coexistence region is characterized by both \eref{eq:fase_LD} and \eref{eq:fase_HD}, expressing the absence of both left and right boundary layers, and respectively associated with pure LD and HD phases. 
Such equations, in combination with \eref{eq:definizione_lambda} and \eref{eq:kirchhoff}, allow us to determine $\rho_+$, $\rho_-$, $\rho_1$ and $\lambda$. 
Note that equation \eref{eq:relazione_bordi} no longer holds in the case of phase separation, as it is based on the assumption of a density profile without shocks, so that we still have 4 unknowns and 4 equations.  
Still in the MF continuum picture, the shock is localized at a position ${x = x_\mathrm{s}}$, where the density jumps from ${\rho_\mathrm{s} < 1/2}$ to ${1 - \rho_\mathrm{s} > 1/2}$, so that the current is continuous (a consequence of equation \eref{eq:NESSbulk} with ${r \to 0}$ in the thermodynamic limit). 
The MF stationary density profile can then be written as \eref{eq:fase_LD_prof}
for ${x \in (0,x_\mathrm{s})}$ (i.e. in the LD portion) and as \eref{eq:fase_HD_prof}
for ${x \in (x_\mathrm{s},1)}$  (i.e. in the HD portion). 
Moreover, the domain-wall position $x_\mathrm{s}$ along with the density $\rho_\mathrm{s}$ can be obtained from conditions ${\lim_{x \to x_\mathrm{s}^-} \rho(x) = \rho_\mathrm{s}}$ and ${\lim_{x \to x_\mathrm{s}^+} \rho(x) = 1-\rho_\mathrm{s}}$, which can be rewritten as
\begin{subequations}
\begin{eqnarray}
F(\rho_\mathrm{s}) & = & F(\rho_1) \, \rme^{-\lambda x_\mathrm{s}}
\, ,
\label{eq:LD-HD-left}
\\
F(1-\rho_\mathrm{s}) & = & F(1 - \beta) \, \rme^{\lambda (1-x_\mathrm{s})}
\, .
\label{eq:LD-HD-right}
\end{eqnarray}
\label{eq:LD-HD}
\end{subequations}

\begin{figure}
	\centerline{\includegraphics[width=0.7\textwidth]{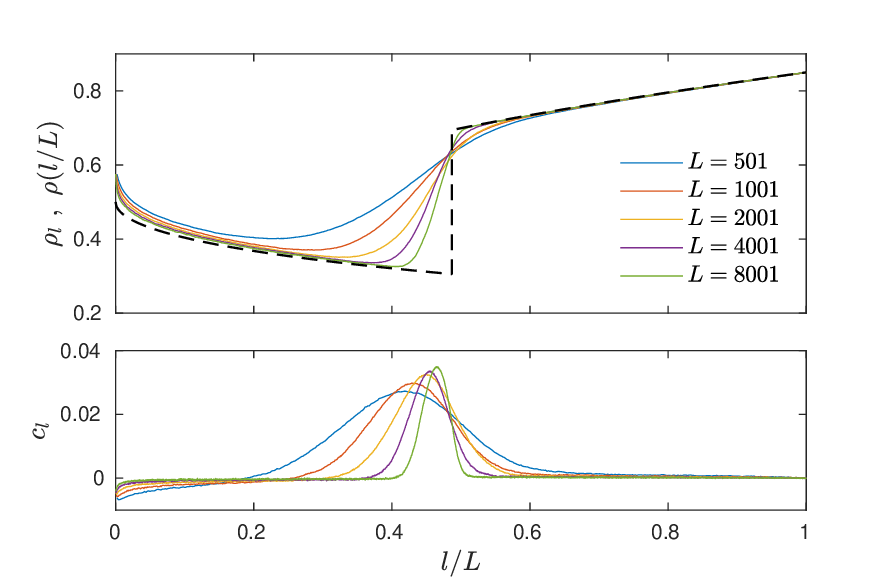}}
	\caption{Same as figure \ref{fig:LD_density} in MC-HD coexistence: ${\alpha = 0.3}$, ${\beta = 0.15}$.
	The black dashed line (MF continuum density profile) is defined piecewise by equations \eref{eq:fase_MC_prof} and \eref{eq:fase_HD_prof}, with the shock position determined by \eref{eq:MC-HD}.}
	\label{fig:MC+HD_density}
\end{figure}

At large injection rate $\alpha$ and intermediate extraction rate $\beta$ we can find MC-HD coexistence, also found in the PBCs case \cite{EPL21}. 
The stationary density profile and the NN covariances, illustrated in figure \ref{fig:MC+HD_density} for ${\alpha = 0.3}$, ${\beta = 0.15}$, exhibit a MC portion on the left (with a left boundary layer) and a HD portion on the right, separated by a domain wall. 
The left boundary layer exhibits qualitatively the same properties as in the MC phase, whereas the value of $\rho_1$ still seems to tend to the MF result in the thermodynamic limit (see figure \ref{fig:rho1}). 
At the MF level the similarity between the LD and MC phases carries over to the corresponding coexistences with the HD phase. 
As a consequence, the density profile is characterized by \eref{eq:fase_MC_prof} and \eref{eq:fase_HD_prof}, respectively associated with pure MC and HD phases. 
Moreover, the conditions determining the domain-wall position $x_\mathrm{s}$ and the corresponding density $\rho_\mathrm{s}$ become
\begin{subequations}
\begin{eqnarray}
F(\rho_\mathrm{s}) & = & F({\textstyle \frac{1}{2}}) \, \rme^{-\lambda x_\mathrm{s}} 
\, ,
\label{eq:MC-HD-left}
\\
F(1-\rho_\mathrm{s}) & = & F(1 - \beta) \, \rme^{\lambda (1-x_\mathrm{s})}
\, .
\label{eq:MC-HD-right}
\end{eqnarray}
\label{eq:MC-HD}
\end{subequations}

The presence of coexisting phases (with a steady domain wall) in this model can be ascribed to non-uniform current profiles, that is, the same phenomenon occurring in the TASEP-LK model, and more specifically in the one with detachment-only Langmuir kinetics, as mentioned in section~\ref{sec:model}. 
Therefore, this is caused by resetting through the inhomogeneities that it induces in the system. 
In spite of the similarities between the LD-HD and MC-HD coexistences, we have noticed that only the latter occurs also in the TASEP-LR with PBCs, whereas the former is a novel feature appearing with OBCs. 
This fact can be roughly explained by considering that, physically, PBCs means joining the two ends of the system. 
Now, it is not possible to join the two ends of a LD-HD density profile like that of figure~\ref{fig:LD+HD_density} without giving rise to a discontinuity, since the LD phase is constrained to have ${\rho < 1/2}$, while the HD phase is constrained to ${\rho > 1/2}$. 
Such a discontinuity would be unstable, as it can be argued from the fact that the LD phase is unable to sustain a ``left'' boundary layer, as well as the HD phase is unable to sustain a ``right'' boundary layer. 
It can be seen that this is not the case for MC-HD coexistence, since the MC phase itself exhibits a ``left'' boundary layer, reaching density ${\rho > 1/2}$, so that a density profile like that of figure~\ref{fig:MC+HD_density} can be closed without inserting extra discontinuities.

\subsection{Phase diagram}
\label{subsec:phd}

\begin{figure}
	\centerline{\includegraphics[width=0.7\textwidth]{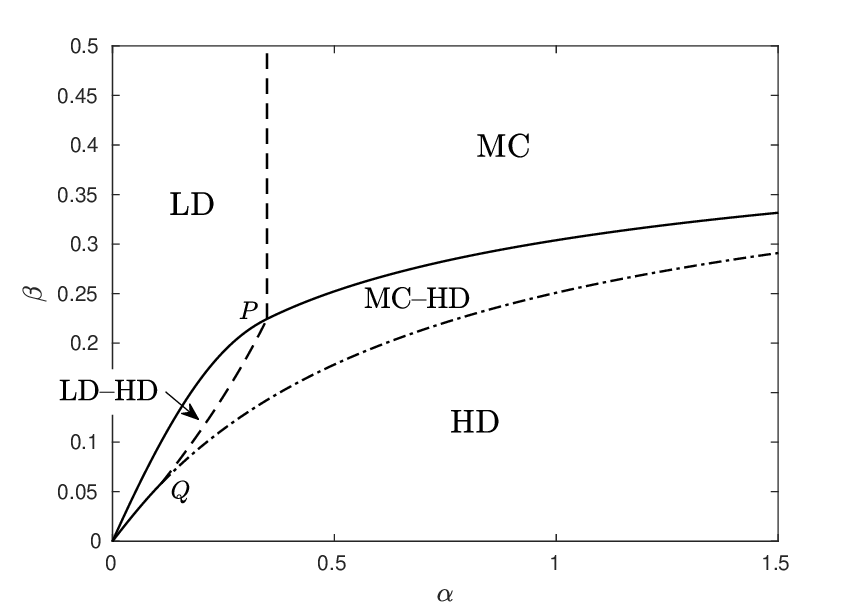}}
	\caption{Stationary state phase diagram for ${R = 0.5}$. 
		The continuous transition between the LD and MC phases is denoted by a dashed line, both for the pure phases and in the region of coexistence with the HD phase. 
		The borders of the coexistence region are denoted by solid lines, except the special one, characterized by a vanishing amplitude of the domain wall, denoted by a dash-dotted line. 
		See the main text for more details.}
	\label{fig:phase_diagram}
\end{figure}

Building on the above characterization of phases and phase coexistences, that can be observed in the stationary state, we can now discuss the related phase diagram. 
The phase diagram is presented in the $(\alpha,\beta)$ plane (see figure \ref{fig:phase_diagram}), fixing the resetting parameter at ${R = 0.5}$, as done in the previous discussion. 
The effect of changes in the latter parameter will be briefly addressed at the end. 
All the phase boundaries can be determined analytically at the MF level. 
Before giving the analytical details, let us only note that, at the MF level, the model with PBCs corresponds to condition ${\beta = 1 - \alpha}$, represented by a straight line in the $(\alpha,\beta)$ phase diagram (the same relation holds for pure TASEP). 
From figure~\ref{fig:phase_diagram} one can see that this line never intercepts the LD-HD coexistence region, which provides a further argument for the absence of this type of coexistence in the PBCs case. 
Let us now analyze all the transition lines appearing in figure~\ref{fig:phase_diagram} one by one. 

The LD and MC phases are separated by a continuous transition, denoted by a dashed line. 
Comparing equations \eref{eq:fase_LD} and \eref{eq:fase_MC}, at this transition we must have ${\rho_+ = \rho_1 = 1/2}$. 
In order to turn this into a condition on the model parameters, we first use equations \eref{eq:relazione_bordi} and \eref{eq:definizione_lambda}, and obtain
\begin{equation}
F(\rho_-) 
= F({\textstyle \frac{1}{2}}) \, \rme^{-R/2}
= {\textstyle \frac{1}{2}} \, \rme^{-(1+R/2)}
\, .
\end{equation}
Taking the inverse of $F$ via \eref{eq:funzione_inversa} (low-density case), we get 
\begin{equation}
\rho_-(R)
= - {\textstyle \frac{1}{2}} \, W_0 \left(-\rme^{-(1+R/2)}\right)
\, .
\label{eq:rhomeno_LD-MC}
\end{equation}
The injection rate, from equation \eref{eq:kirchhoff}, is then found as
\begin{equation}
\alpha 
= 2 \rho_-(R) \left[ 1-\rho_-(R) \right]
\, ,
\label{eq:alpha_LD-MC}
\end{equation}
where $\rho_-(R)$ is given by equation \eref{eq:rhomeno_LD-MC}. 
As a consequence, the critical value of $\alpha$ is independent of $\beta$ and turns out to be a decreasing function of $R$, taking value $1/2$ at ${R = 0}$ (pure TASEP) and vanishing in the limit ${R \to \infty}$.

The HD phase is separated from the LD and MC phases by the corresponding coexistence regions. 
The boundaries of these regions in the phase diagram, denoted in figure \ref{fig:phase_diagram} by solid or dash-dotted lines, can be obtained by imposing the conditions that the domain wall reaches one end of the lattice, that is by setting ${x_\mathrm{s} = 0}$ or ${x_\mathrm{s} = 1}$ in equations \eref{eq:LD-HD} and \eref{eq:MC-HD}. 
In all these cases we have ${\rho_- = 1 - \beta}$, so the corresponding lines in the phase diagram are obtained from equation \eref{eq:kirchhoff} as
\begin{equation}
\alpha = \frac{\beta (1-\beta)}{1-\rho_1(\beta,R)}
\, ,
\label{eq:alfadibeta}
\end{equation}
where the function $\rho_1(\beta,R)$ depends on the boundary one considers. 
For instance, the boundary between the LD-HD coexistence and the LD pure phase is obtained by setting ${x_\mathrm{s} = 1}$ in equations \eref{eq:LD-HD}, that is by imposing that the HD portion of the LD-HD coexistence disappears. 
Recalling \eref{eq:definizione_lambda}, this yields
\begin{equation}
F(\beta) 
= F(\rho_1) \, \rme^{-R(1-\rho_1)}
= \frac{\rme^{-R}}{1-R/2} \, F\big((1-R/2)\rho_1\big)
\, ,
\label{eq:transizione_LD-(LD+HD)}
\end{equation}
which can be solved for $\rho_1$ by means of \eref{eq:funzione_inversa} (low-density case), giving
\begin{equation}
\rho_1(\beta,R) 
= -\frac{1}{2-R} \, W_0 \left(-(2-R)\beta\,\rme^{-(2\beta-R)}\right)
\, .
\label{eq:rho0_LD-(LD+HD)}
\end{equation}
Similarly, the boundary between the LD-HD coexistence and the HD pure phase is obtained by setting ${x_\mathrm{s} = 0}$ in equations \eref{eq:LD-HD}. 
Still recalling \eref{eq:definizione_lambda}, this yields
\begin{equation}
F(1-\beta) 
= F(1-\rho_1) \, \rme^{-R(1-\rho_1)}
= \frac{1}{1+R/2} \, F\big((1+R/2)(1-\rho_1)\big)
\, ,
\label{eq:transizione_(LD+HD)-HD} 
\end{equation}
and hence by \eref{eq:funzione_inversa} (high-density case)
\begin{equation}
\rho_1(\beta,R) 
= 1 + \frac{1}{2+R} \, W_{-1} \left(-(2+R)(1-\beta)\,\rme^{-2(1-\beta)}\right)
\, .
\label{eq:rho0_(LD+HD)-HD}
\end{equation}
Finally, for the boundaries of the MC-HD coexistence we set ${x_\mathrm{s} = 1}$ (MC pure phase) or  ${x_\mathrm{s} = 0}$ (HD pure phase) in equations \eref{eq:MC-HD}. 
Still taking into account \eref{eq:definizione_lambda}, we obtain respectively 
\begin{equation}
\rho_1(\beta,R)
= 1 - \frac{1}{R}
\,\big[ 2\beta - 1 - \ln(2\beta) \big]  
\, ,
\label{eq:rho0_MC-(MC+HD)}
\end{equation}
or 
\begin{equation}
\rho_1(\beta,R)
= 1 - \frac{1}{R}
\,\big[ 1 - 2\beta - \ln(2-2\beta) \big]  
\, .
\label{eq:rho0_(MC+HD)-HD} 
\end{equation}
The latter boundary is the only one characterized by a vanishing height of the domain wall (${\rho_\mathrm{s} = 1 - \rho_\mathrm{s} = 1/2}$), and for this reason it is denoted in figure \ref{fig:phase_diagram} by a special (dash-dotted) line.
A similar effect was observed in \cite{ParmeggianiFranoschFrey04} for the TASEP-LK. 
Moreover, we easily see that equations \eref{eq:rho0_MC-(MC+HD)} and \eref{eq:rho0_(MC+HD)-HD} both imply ${\rho_1(1/2,R) = 1}$ for all $R$. 
As a consequence, both lines bounding the MC-HD coexistence regions are characterized by ${\beta \to 1/2}$ for ${\alpha \to \infty}$.

\begin{figure}
	\centerline{\includegraphics[width=0.7\textwidth]{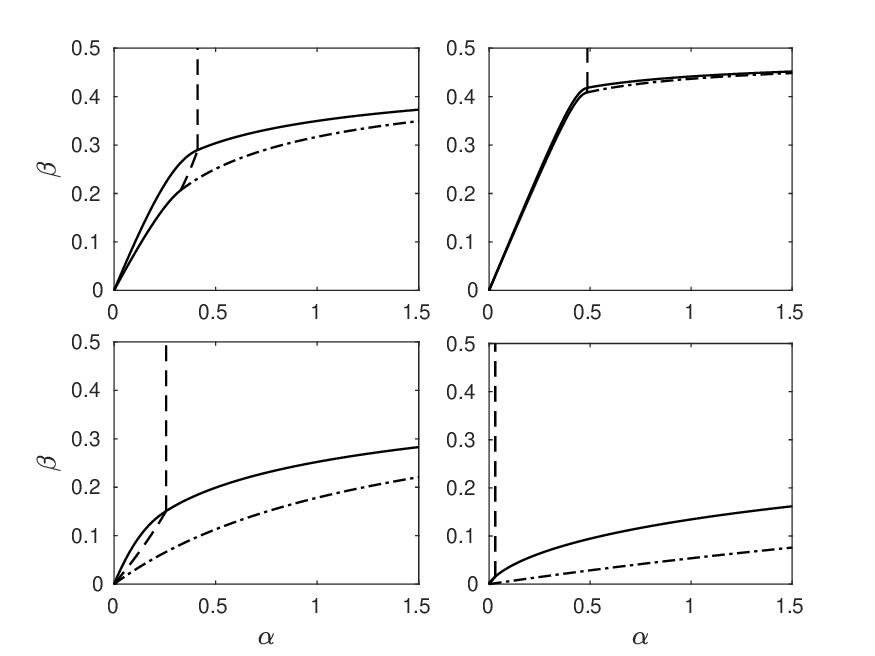}}
	\caption{Stationary state phase diagrams for ${R = 0.25}$ (top left), ${R = 0.03}$ (top right), ${R = 1}$ (bottom left), ${R = 5}$ (bottom right). 
		The different phase regions and transitions can be deduced from figure~\ref{fig:phase_diagram}.}
	\label{fig:phase_diagram_multi}
\end{figure}

In the end, let us consider the boundary between the two coexistence regions, namely LD-HD and MC-HD. 
Such regions are separated by a continuous transition (dotted line in figure \ref{fig:phase_diagram}), which can be obtained by observing that it is characterized by ${\rho_+ = 1/2}$ (as for the continuous transition between the LD and MC pure phases) and ${\rho_- = 1 - \beta}$ (which holds in both regions). 
Plugging both conditions into equation \eref{eq:kirchhoff}, we obtain
\begin{equation}
\alpha 
= 2 \beta (1-\beta) 
\, ,
\label{eq:transizione_(LD+HD)-(MC+HD)}
\end{equation}
independent of $R$. 
The end points of this line, labeled $P$ and $Q$ in figure \ref{fig:phase_diagram}, can be obtained by setting ${\rho_1 = 1/2}$ in equations \eref{eq:transizione_LD-(LD+HD)} and \eref{eq:transizione_(LD+HD)-HD} respectively, which yields 
\begin{equation}
\beta
= - {\textstyle \frac{1}{2}} W_0 \left(-\rme^{-(1+R/2)}\right)
\, 
\end{equation}
for point $P$ and
\begin{equation}
\beta
= 1 + {\textstyle \frac{1}{2}} W_{-1} \left(-\rme^{-(1+R/2)}\right)
\,
\label{eq:betaQ}
\end{equation}
for point $Q$. 
The corresponding values of the injection rate $\alpha$ can of course be obtained from equation \eref{eq:transizione_(LD+HD)-(MC+HD)}. 
Notice that the right hand side of equation \eref{eq:betaQ} is positive only if ${R < R_0 = 2 \, (1 - \ln 2) \simeq 0.614}$, which implies that point $Q$ and the boundary between the LD-HD coexistence and the HD pure phase disappear for ${R > R_0}$.
In figure~\ref{fig:phase_diagram_multi} we report a sequence of $(\alpha,\beta)$ planes of the phase diagram, for different fixed values of $R$, showing in particular the change of topology occurring for ${R > R_0}$. 
We can see also that, when $R$ gets smaller and smaller, then the phase diagram tends continuously to that of pure TASEP. 

The stationary-state phase diagram we have obtained relies on estimates of the local densities $\rho_1$ and $\rho_\pm$ and on the balance equation for the current, equation \eref{eq:kirchhoff}, whose right-hand side is based on a MF assumption. 
On the other hand, we have got considerable numerical evidences that the values of $\rho_1$ and $\rho_\pm$, computed by the MF theory, may be exact in the thermodynamic limit, as well as equation \eref{eq:kirchhoff}. 
As a consequence, we can expect that the phase diagram presented here is also exact, as it happens for the MF stationary-state phase diagram in the case of pure TASEP \cite{Derrida98,Schutz,Evans2007,ZiaReview,TransportBook} and, at the level of numerical evidence, in some variants such as the TASEP-LK \cite{ParmeggianiFranoschFrey04}. 
Conversely, let us note that certain evidences collected from \cite{EPL21} suggest that this is likely not to be the case in the current model with PBCs, due in particular to a breakdown of equations \eref{eq:fase_LD} and \eref{eq:fase_HD}, i.e. to the onset of (very small but measurable) extra boundary layers in LD and HD phases. 

\begin{figure}
	\centerline{\includegraphics[width=0.7\textwidth]{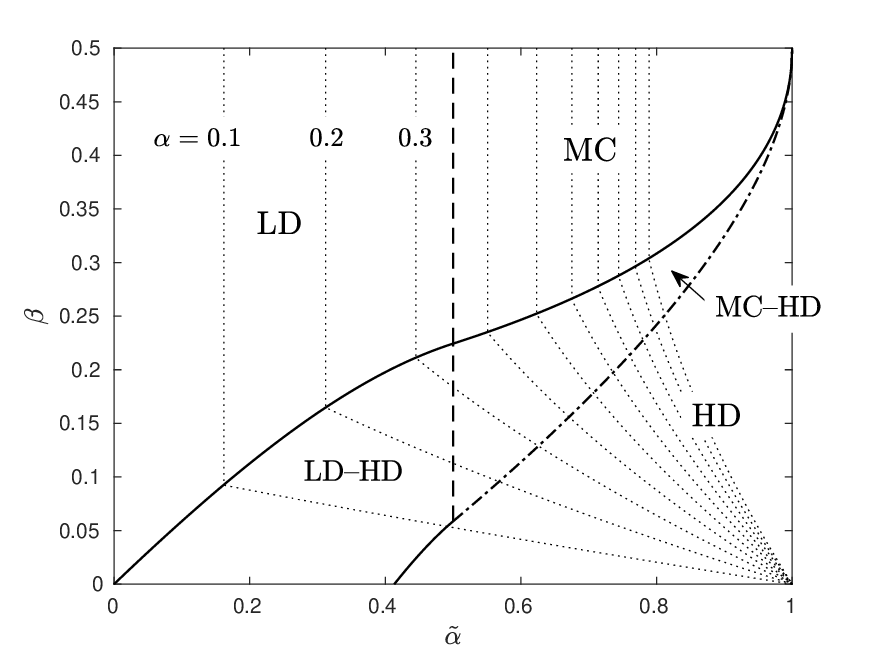}}
	\caption{Stationary state phase diagram of detachment-only TASEP-LK, projected on the plane defined by equation \eref{eq:mapping_1}, for ${R = 0.5}$. 
		Thick lines denote phase transitions (line coding and phase labels as in figure~\ref{fig:phase_diagram}). 
		Thin dotted lines denote constant-$\alpha$ lines for the corresponding TASEP-LR ($\alpha$ values equally spaced from $0.1$ to $1.0$).
		See the main text for more details.}
	\label{fig:phase_diagram_cut}
\end{figure}

Let us also briefly return to the connection, mentioned in section~\ref{sec:model}, between the current model and TASEP-LK with detachment-only Langmuir kinetics. 
The MF phase diagram of the latter model, worked out analytically in \cite{BonninKernYoungStansfieldRomano2017}, exhibits two main differences with respect to that of TASEP-LR. 
First, the transition line between LD and MC (both as pure phases and in coexistence with HD) is found at a constant injection rate ${\tilde{\alpha} = 1/2}$.
Moreover, there exists a threshold value of the detachment rate ${\Omega_{\mathrm{D},0} = 1 - \ln 2}$, above which the pure HD phase disappears completely.  
According to our previous discussion, the ``effective'' TASEP-LK must be characterized by 
\begin{equation} 
\Omega_\mathrm{D} = R \left( 1-\tilde{\alpha} \right)
\, ,
\label{eq:mapping_1} 
\end{equation} 
which physically means that local resetting introduces a trade-off between particle detachment (from the bulk of the system) and hopping from the resetting node to the bulk (i.e.~the ``effective injection'' of TASEP-LK).
In the appropriate 3-dimensional parameter space $(\tilde{\alpha},\beta,\Omega_\mathrm{D})$, equation~\eref{eq:mapping_1} represents an oblique plane, with a slope defined by the resetting rate $R$. 
Using the analytical results in \cite{BonninKernYoungStansfieldRomano2017}, we can draw a projection of the phase diagram of (detachment-only) TASEP-LK onto this plane. 
The result is displayed (as thick lines) in figure~\ref{fig:phase_diagram_cut}, for the usual value ${R = 0.5}$. 
Of course, this is not yet the phase diagram of TASEP-LR, but nonetheless meaningful, in particular because it shows that the persistence of the pure HD phase in TASEP-LR (with no threshold, at odds with TASEP-LK) is to be ascribed precisely to the trade-off mechanism represented by~\eref{eq:mapping_1}.\footnote{As $\tilde{\alpha}$ moves from $0$ to $1$, $\Omega_\mathrm{D}$ moves from $R$ to $0$, so that it necessarily gets below the threshold value $\Omega_{\mathrm{D},0}$. We note in particular that ${\Omega_{\mathrm{D},0} = R_0/2}$, which corresponds to ${\tilde{\alpha} = 1/2}$. As a consequence, the disappearance of the pure HD phase in TASEP-LK corresponds to the topological change in the TASEP-LR phase diagram.} 
To complete the mapping, we still have to determine $\tilde{\alpha}$ as a function of the real control parameters of TASEP-LR, that is $\alpha$, $\beta$ and $R$. 
In particular we have to take into account the balance equation~\eref{eq:kirchhoff}, with ${\rho_1 = \tilde{\alpha}}$. 
In the whole region where the HD phase is present (both as a pure phase and in coexistence with LD or MC) we have ${\rho_- = 1-\beta}$, and the resulting equation simplifies (also being independent of $R$) as 
\begin{equation} 
\tilde{\alpha} = 1 - \frac{\beta (1-\beta)}{\alpha} 
\, .
\label{eq:mapping_2} 
\end{equation} 
Conversely, in the region of pure LD and MC phases, we know that $\rho_-$ no longer depends on $\beta$, so that also $\tilde{\alpha}$ depends only on $\alpha$ and $R$. 
As a consequence, constant-$\alpha$ lines, that we report in figure~\ref{fig:phase_diagram_cut}, exhibit a kink at the intersection with the boundary between the aforementioned regions, and appear as vertical straight lines in the latter. 
These lines physically represent the role of $\alpha$ in controlling the resetting process, and help explain some other peculiarities of the phase diagram of TASEP-LR. 
In particular, the reason why the transition between LD and MC no longer occurs at constant injection rate (at odds with TASEP-LK) is to be ascribed precisely to the nonlinear (actually even singular) relation between $\alpha$ and $\tilde{\alpha}$.

\subsection{Small and large resetting}

\begin{figure}
	\centerline{\includegraphics[width=0.7\textwidth]{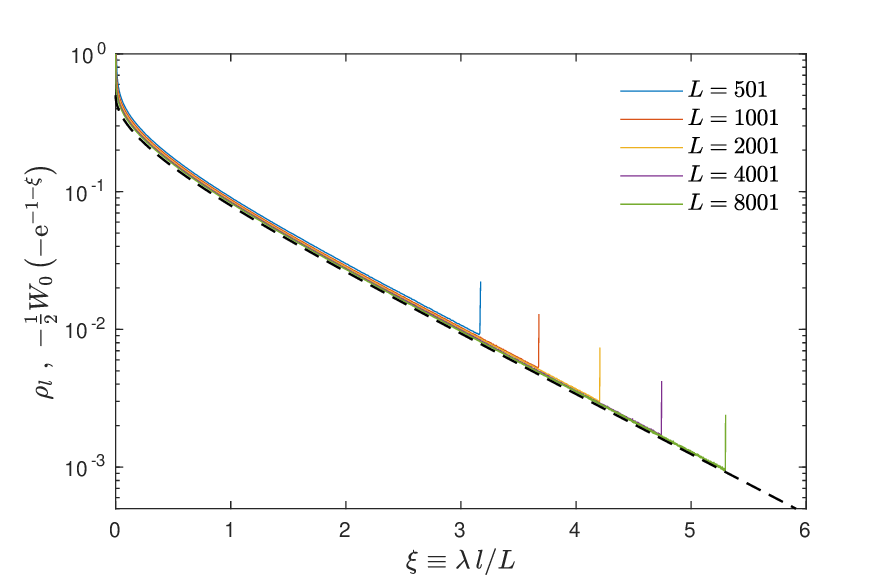}}
	\caption{Stationary density profile as a function of the scaled position variable ${\xi \equiv \lambda \, l/L}$, in the large resetting regime: ${r = R/L = 0.5}$, ${\alpha = 0.7}$, ${\beta = 0.4}$. 
	Colored lines denote KMC results. 
	A black dashed line denotes the MF continuum profile, corresponding to equation \eref{eq:fase_MC_prof} with ${\lambda x = \xi}$, i.e.\ the function $-\frac{1}{2} W_0\left(-\rme^{-1-\xi}\right)$.}
	\label{fig:LargeResetting}
\end{figure}

We conclude this section by briefly reporting the properties of the stationary states in the small and large resetting regimes. 
In the small resetting regime, we have ${R \to 0}$ in the thermodynamic limit, i.e.~the resetting rate $r$ vanishes faster than $1/L$, and it is easy to check that the stationary state trivially reduces to that of pure TASEP, unperturbed by resetting.
In the large resetting regime we have ${R \to \infty}$ and resetting dominates over injection and extraction processes. 
In the latter case, both MF and KMC results at finite size $L$ exhibit a stationary density profile characterized by ${\rho_1 \to 1}$ and ${\rho_L \to 0}$ for increasingly large $L$. 
The approach to the thermodynamic limit is slow, but numerical inspection suggests to assume that the bulk density profile for ${L \to \infty}$ is MC-like, i.e. that it can be described by equation \eref{eq:fase_MC_prof}, even though with ${\lambda \to \infty}$. 
This assumption is corroborated by comparing finite-size KMC profiles with the analytical profile \eref{eq:fase_MC_prof}, as a function of a scaled position variable ${\xi \equiv \lambda \, l/L}$, where the finite-size value of $\lambda$ is computed through equation \eref{eq:definizione_lambda}. 
The results are displayed in figure \ref{fig:LargeResetting}, for a specific parameter choice. 
Our ansatz can also be rationalized by observing that, in the intermediate resetting regime, taking the limit ${R \to \infty}$, the MC phase ``invades'' the whole phase diagram, as it can be argued from figure~\ref{fig:phase_diagram_multi} and, analytically, from the discussion in the previous subsection. 
In particular, equations \eref{eq:rhomeno_LD-MC} and \eref{eq:alpha_LD-MC} entail ${\alpha \to 0}$ at the continuous transition between the MC and LD phases, whereas equations \eref{eq:rho0_MC-(MC+HD)} and \eref{eq:alfadibeta} entail ${\alpha \to \infty}$ at the boundary between the MC phase and the MC-HD coexistence, for any fixed ${\beta > 0}$. 

\begin{figure} 
	\centerline{\includegraphics[width=0.7\textwidth]{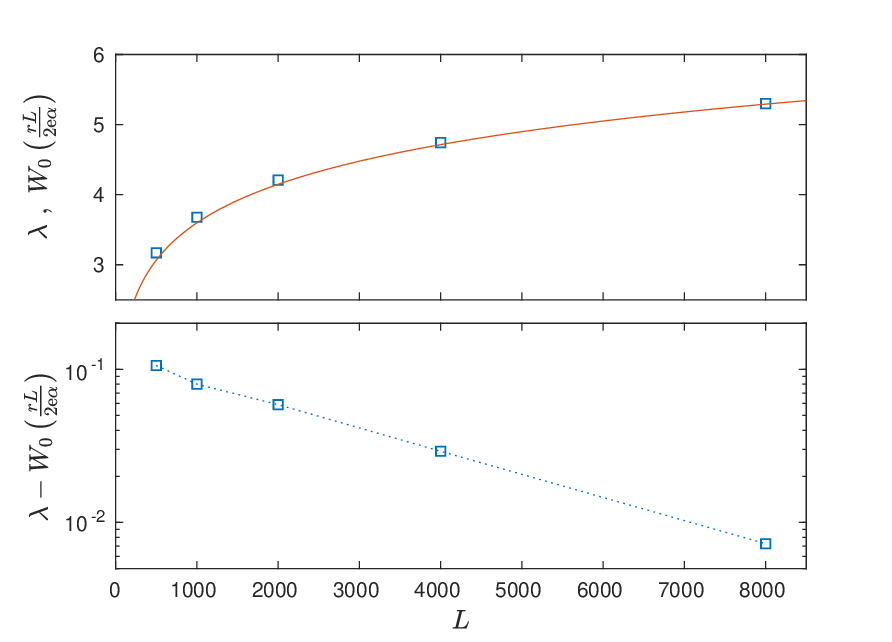}}
	\caption{Inverse characteristic length $\lambda$ of the bulk density profile as a function of the lattice size $L$, in the large resetting regime: ${r = R/L = 0.5}$, ${\alpha = 0.7}$, ${\beta = 0.4}$. 
	Top panel: finite-size $\lambda$ from KMC (blue squares) and the asymptotic expression \eref{eq:lambda}, i.e.\ $W_0 \left( \frac{rL}{2 \rme \alpha} \right)$ (red line). 
	Bottom panel: difference (the dotted line is a guide for the eye).}
	\label{fig:lambdaL}
\end{figure}

In order to understand how $\lambda$ diverges for ${L \to \infty}$, we first use equation \eref{eq:definizione_lambda} in combination with the (exact) balance equation for the currents, ${\alpha (1-\rho_1) = \beta \rho_L}$, obtaining
\begin{equation} 
\lambda = \frac{R}{\alpha} \, \beta \rho_L
\, .
\label{eq:lambda_1}
\end{equation} 
Then, keeping in mind that, for large $L$, the density near the right boundary is vanishingly small, from the discrete stationary-state MF equations \eref{eq:NESSL} and \eref{eq:NESSbulk} we get ${\beta \rho_L \simeq \rho_{L-1} \simeq \rho_{L-2} \simeq \dots}$ (at the leading order, and within a finite distance from the right boundary). 
Moreover, with an eye to figure \ref{fig:LargeResetting}, we argue that in this microscopic region (except the rightmost node $L$, affected by the extraction process) the density can be approximated as well by the continuum expression \eref{eq:fase_MC_prof} evaluated for ${x = 1}$. 
Also taking into account that the Lambert function in \eref{eq:fase_MC_prof} can be replaced by its argument, being itself vanishingly small, we obtain
\begin{equation} 
\beta \rho_L \simeq \textstyle{\frac{1}{2}} \, \rme^{-1-\lambda}
\, .
\label{eq:lambda_2}
\end{equation} 
The latter equation along with \eref{eq:lambda_1} provide an equation for $\lambda$, which can be solved once again by means of a Lambert function, yielding
\begin{equation}
\lambda \simeq W_0 \left( \frac{R}{2 \rme \alpha} \right) 
\, .
\label{eq:lambda}
\end{equation}
As a result, we see that $\lambda$ grows logarithmically with respect to $R$, as one can argue from the asymptotic expansion ${W_0(z) = \ln z - \ln \ln z + o(1)}$ for ${z \to \infty}$. 
In figure \ref{fig:lambdaL} we plot, as a function of $L$ (and with the same parameters used for figure \ref{fig:LargeResetting}), the finite-size $\lambda$ obtained from KMC simulations (still through $\rho_1$ and equation \eref{eq:definizione_lambda}), along with the analytical expression \eref{eq:lambda}, and their difference. 
These last results suggest that the asymptotic behaviour of $\lambda$ expressed by equation \eref{eq:lambda} may be exact, and more generally that, even in the large resetting regime, the MF theory may describe exactly the bulk behavior of the system in the thermodynamic limit.

\section{Conclusions} 
\label{sec:discuss}

We have studied the stationary state of a TASEP-LR with OBCs and resetting at the injection node, using the MF approximation and KMC simulations, which agree remarkably well in the thermodynamic limit. 
As in the case with PBCs, a relationship with TASEP-LK is established and three regimes can be identified, depending on how the resetting rate scales with the lattice size in the thermodynamic limit. 
The most interesting case is the intermediate resetting regime, where we have characterized the pure phases and the phase separation phenomena, discussing also their boundary layers, and mapped out analytically the phase diagram at the MF level, which we conjecture to be exact. 
With respect to the PBCs case, an additional phase separation phenomenon is found, namely coexistence between LD and HD phases, which we argue to induce a change of topology in the phase diagram, depending on the value of the macroscopic resetting rate $R$. 
In the large resetting regime we have also characterized the scaling of the bulk density profile with the lattice size. 

Regarding possible applications of the model to real (biological) systems, in the context of the polyribosome the TASEP-LR may be viewed as an effective way of incorporating, into the classical TASEP description of ribosome dynamics~\cite{MacDonaldGibbsPipkin1968}, the drop-off effect (i.e. the premature termination of the translation process, due to stalled ribosomes)~\cite{BonninKernYoungStansfieldRomano2017}, along with so-called ribosome rescue and recycling~\cite{FranckenbergBeckerBeckmann2012}. 
Actually, it has to be noted that most recent studies support the idea that cells are able to monitor ribosome collisions, and dynamically tune the initiation rate (i.e.~the TASEP $\alpha$ parameter) in such a way to avoid excessive ribosome density (see \cite{NeelagandanLambertiCarvalhoGobetNaef2020} and references therein). 
According to this paradigm, only a very small portion of the phase diagram at low~$\alpha$ (mostly the LD phase) could be accessible to experiments. 
As a consequence, we argue that possible experiments aimed at observing the phase transition and coexistence scenario, predicted by the model, could in principle be realized through a substantial reduction of the termination rate $\beta$, by acting on the specific molecular factors involved in that process. 
On the other hand, some theoretical works~\cite{BonninKernYoungStansfieldRomano2017,CiandriniStansfieldRomano2013} have speculated about the importance of analyzing models even in conditions other than those experimentally observable. 
In particular, the cited papers adjust the TASEP hopping rates, according to an established criterion, in order to obtain gene-specific models. 
Then, performing a genome-wide analysis, they demonstrate (for \emph{Saccaromyces cerevisiae}) that mRNAs coding for proteins involved in different types of biological functions are significantly correlated with different phase behaviors, that emerge in the corresponding models upon artificially increasing the injection rate $\alpha$. 
Of course, in this respect our model with homogeneous hopping rates lacks specificity, but in principle it may serve as a basis for more detailed investigations, along the lines of \cite{BonninKernYoungStansfieldRomano2017,CiandriniStansfieldRomano2013}. 

From the theoretical point of view, our results suggest several possible future developments, and work is in progress along some of these lines. 
The issue of possible exactness of the MF phase diagram (which we believe to be a peculiarity of OBCs) is certainly worth a more rigorous investigation. 
Models with additional interactions \cite{Katz1984,Antal2000,Dierl2012,Dierl2013} or different geometries (resetting at a node different from the injection node, or to multiple nodes \cite{Boyer}) may also be investigated. 
Another line of research might consider the relaxation process towards the stationary state, in which dynamical transitions (not corresponding to any static transition) have been pointed out in (T)ASEP \cite{deGierEssler05,deGierEssler08,ProemeBlytheEvans11} and (at least with approximate methods), TASEP-LK \cite{Botto2019,Botto2020} and exclusion processes with interactions \cite{Botto2018,Puccioni}. 
We also hope that these and other theoretical results on local resetting will stimulate advancements in ad hoc experimental studies, which to the best of our knowledge have been so far limited to resetting in single-particle systems \cite{Roichman,Ciliberto}.





\end{document}